\documentstyle[11pt,paspconf,epsf]{article}

\begin{document}

\title{Large-scale Motions in the Universe:
    Observations and Simulations}

\author{D. H. Gudehus}
\affil{Physics and Astronomy Department, Georgia State University,
    Atlanta, GA 30303}



\begin{abstract}
The peculiar radial velocities of eight clusters of galaxies, as 
determined from a variety of distance indicators, are consistent with a
bulk flow not greatly different in direction and magnitude from the
velocity vector of the Local Group relative to the\break microwave
background radiation.  The bulk flow vector is directed to $l=266\deg
\pm 18\deg, b=17\deg \pm 15\deg$ with a magnitude of $788\pm
113$ {\hbox{km\ s$^{-1}$}}.  N-body simulations of gravitationally induced
peculiar motions from a set of over 1000 clusters of  known redshift give
reasonably good agreement for the peculiar velocities of the  Local Group
and Virgo both in magnitude  and direction.  Most of the accelerating 
mass lies within about  40$h^{-1}$ Mpc but more distant clusters are
necessary  to fully  account for the magnitude and direction.
\end{abstract}




\section{Introduction}

It has been more than 25 years since the first peculiar velocity of a cluster of
galaxies was reported (Gudehus 1973), more than 20 years since peculiar velocities
were first referenced to the microwave background radiation (MBR) (Gudehus
1978), accurately measured for the first time by Smoot, Gorenstein, and Muller
(1977), and more than 10 years since Dressler {\it et al.} (1987) first reported
that galaxian peculiar velocities were correlated in direction.  The first part of
this time period, however, was marked by the fixed idea that the universal expansion
was ``quiet'' (Visvanathan and Sandage 1977 and references
therein; Tammann 1984).  Tammann (1984), for example, claimed that my 1978 peculiar
velocity for Virgo was ``impossible in size {\it and} sign''.  These early contrary
claims have been explained (Gudehus 1989b, 1991) as due to {\it i)} a neglect to
reference redshifts to the MBR frame and {\it ii)} a variety of systematic
effects.  The  current view is that peculiar velocities exist at some level out to
15,000  {\hbox{km\ s$^{-1}$}},  but with some observers finding significantly
smaller values at greater distance (Riess {\it et al.} 1995;  Giovanelli {\it et
al.} 1998) than others  (Gudehus 1989b;  Lucey {\it el al.} 1991;  Bothun {\it et
al.} 1992;  Mould {\it et al.} 1993; Hudson {\it et al.} 1999);  Willick (1999),
and one group (Lauer \& Postman 1994) finding a significantly different vector
direction for the most distant clusters.

In this paper I will use relative distance moduli taken from my previous studies,
solve for a preferred set of revised cluster peculiar velocities, and compute an
average direction for the cosmic flow.  I will then, from an N-body simulation 
based on masses assigned to clusters of known redshift, calculate predicted peculiar
velocities for the LG and Virgo.  

\section{The Data}

The data sets used in this study comprise $m$* magnitudes (Gudehus 1989b) for ten
clusters, reduced galaxian radius parameters, $r_g$, (Gudehus 1991) for four
clusters, and relative magnitudes  of seventeen selected clusters computed as
weighted averages from several independent distance indicators (Gudehus 1995)
(including first-ranked magnitude and richness-corrected nuclear magnitudes).  The
$m$* and $r_g$ data were combined and fitted by nonlinear least squares to the
redshift-magnitude diagram with $q_0$ assumed to be 0.5, luminosity evolution assumed
to be 0.75 magnitudes per unit $z$, and all redshifts corrected to the Cosmic
Background Explorer (COBE) results for the LG's motion relative to the MBR (Smoot
{\it et al.} 1991).  The nearest five clusters were given zero weight in the fit
since their peculiar velocities produce substantial deviations from the curve and are
nonrandomly distributed.  The standard deviation about the curve fitted to the
remaining points is 0.064 mag.  Peculiar velocities were derived from the selected
clusters in a similar way and include three duplicates from the first set, i.e.,
Virgo, Fornax, and Hydra I, which were given zero weight in the fit.  Although the
standard deviation about the fit is larger, i.e., 0.11 mag, than for the first set,
the preferred peculiar velocities of the duplicates are taken from the second set
since in that set each cluster's point represents a weighted average of several
independent distance indicators (three for Fornax and Hydra I, and six for Virgo). 
Table~\ref{tbl-1} lists for each cluster, the  peculiar velocity observations.  
Column 1 gives the identity, column 2 the redshift velocity relative to the MBR,
column 3 the  distance modulus relative to Virgo consistent with the peculiar
velocities, and column 4 the radial peculiar velocity, respectively.  

\begin{table}
\caption{Cluster Peculiar Velocity Data.} \label{tbl-1}
\begin{center}
\begin{tabular}{lcrr}
        & $cz_u$                  &            &  $v_{r,{\rm pec}}\ \ \ $  \\
Cluster & {\hbox{(km\ s$^{-1}$)}} & $\Delta m$ & {\hbox{(km\ s$^{-1}$)}} \\ 
\tableline
Virgo      & 1473  & 0.00  & $ 387\pm 66\ $ \\
Ursa Major & 1140  & 0.25  & $ -79\pm $111  \\
Fornax     & 1428  & 0.26  & $ 202\pm 78\ $ \\
Hydra I    & 4027  & 2.44  & $ 665\pm$167   \\
Perseus    & 5179  & 3.62  & $-541\pm$399   \\
A1367      & 6880  & 3.71  & $ 883\pm$200   \\
Coma       & 7185  & 4.06  & $ 165\pm$214   \\
A2199      & 9081  & 4.71  & $-335\pm$297   \\
\end{tabular}
\end{center}
\end{table}

\begin{figure}
\plotfiddle{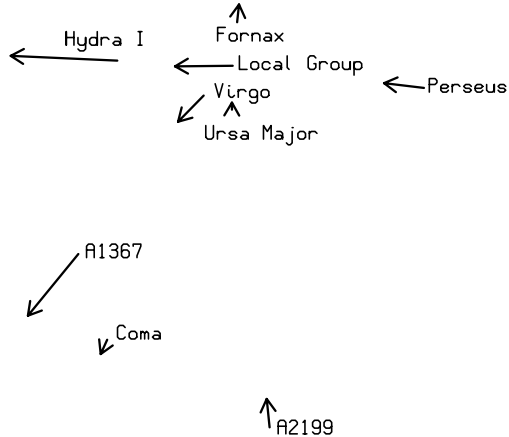}{2.2in}{0}{60}{60}{-220}{-10}
\caption{Peculiar radial velocity vectors of eight nearby clusters as
projected on  the best-fitting plane through their positions on the sky.   The
weighted average of the vectors, excluding the  Local Group, is directed towards
$l=272$\deg $\pm$ 18\deg, $b=48$\deg $\pm$ 15\deg, and differs by 19\deg\  from the
direction of the Local Group's motion relative to the MBR.} \label{fig-1}
\end{figure}

Figure~\ref{fig-1} shows the radial peculiar velocity vectors projected on the best fitting
plane passing through the eight clusters.  The pole of this plane is in the 
direction
$\alpha=6\fh7$, $\delta= 16\deg$, or $l=198\deg, b=6\deg$
(1950.0 equinox).  The angle between the pole and the MBR
apex is   $77\deg$. The angle between the pole and the supergalactic pole is
$31\deg$. 

\begin{figure}
\plotfiddle{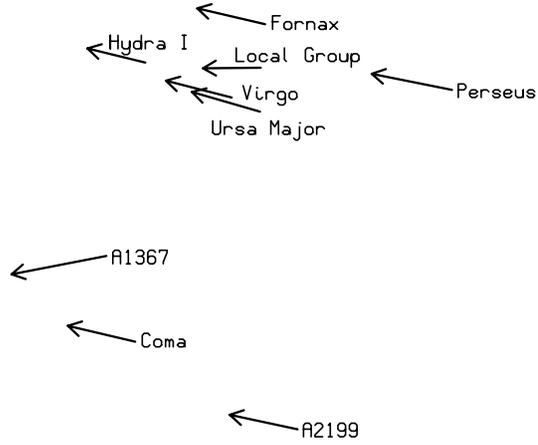}{2.2in}{0}{60}{60}{-220}{-10}
\caption{Peculiar space velocity vectors of eight nearby clusters as
projected on  the best-fitting plane through their positions on the sky.  The 
tangential components are taken from a bulk velocity vector fitted to the radial
components by nonlinear least squares.  The bulk flow vector is directed towards
$l=266$\deg $\pm$ 18\deg, $b=17$\deg $\pm$ 15\deg with a magnitude of
788 $\pm$ 113 {\hbox{km\ s$^{-1}$}}, and differs by
15\deg\  from the direction of the Local Group's motion relative to the MBR.} 
\label{fig-2}
\end{figure}

\begin{figure}
\plotfiddle{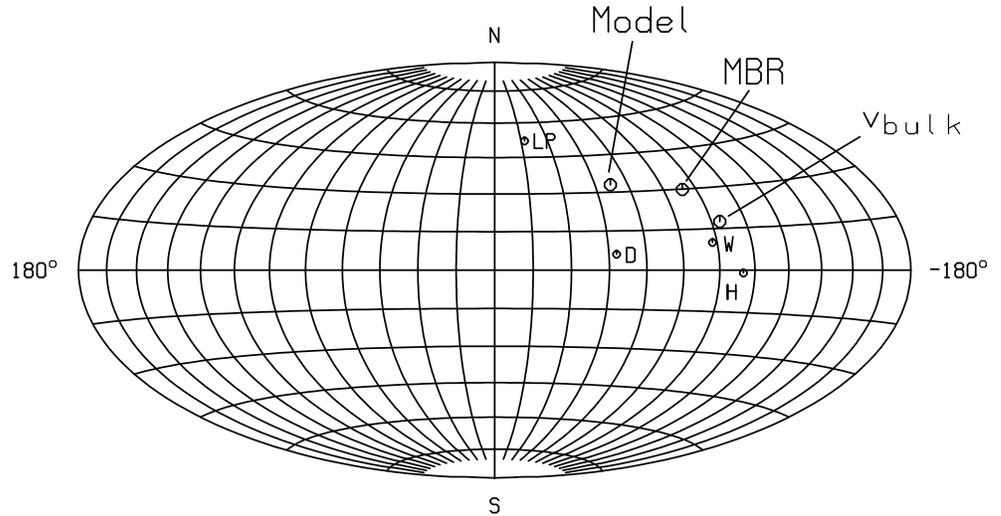}{2.2in}{0}{70}{70}{-190}{-15}
\caption{The directions, on a Hammer-Aitoff projection in galactic 
coordinates, of the observed bulk flow velocity vectors of the
present study and those of Dressler {\it et al.} (1987) (D), Hudson {\it et al.} 
(1999) (H), Lauer and Postman (1994) (LP), and Willick (1999) (W), all with
respect to the microwave background radiation (MBR), whose direction is shown as well. 
Also shown is the direction of the LG's motion as derived from an N-body simulation with
1009 clusters of known redshift.} 
\label{fig-3}
\end{figure}

A bulk flow vector, solved for by nonlinear least squares, is directed to
$l=266\deg \pm 18\deg, b=17\deg \pm 15\deg$ with a magnitude of
$788\pm 113$ {\hbox{km\ s$^{-1}$}}.  The mean error of the fit is 136 
{\hbox{km\ s$^{-1}$}}, and the vector
differs by $15\deg$ from the direction of the LG's motion relative to the MBR.
The average depth is 4549 {\hbox{km\ s$^{-1}$}} and the average weighted depth
is 2214 {\hbox{km\ s$^{-1}$}}.
Bulk flow tangential velocity components added to the observed radial components
give the space vectors plotted in Figure~\ref{fig-2}.
In Figure~\ref{fig-3} I have plotted
on a Hammer-Aitoff projection in galactic coordinates, the directions of the bulk
flow vectors of this study and several others.  

\section{Simulations}

In  the simulations presented here I consider
the gravitational accelerations due to clusters of galaxies for which 
redshifts are presently known.
In the first step of the simulation calculation, clusters within a given redshift
range are selected from a catalogue of all known clusters.  For each cluster the
coordinates and distances based on a given Hubble parameter, $H_0$ and deceleration
parameter, $q_0$, are computed.  The distances are then reduced to values appropriate
for a young universe and the model universe allowed to expand.  At each uniform time
step, the acceleration from each cluster on the LG and Virgo are computed, and the
velocities are computed from the integrals.  Ten free parameters are used in the
simulation.  Besides $H_0$ and $q_0$, there are the mass of a richness 2 cluster, 
${\cal M}_2$; the factor to be applied to the number of galaxies in a richness 0 cluster
in describing a cluster of unknown richness, $F$; the power describing how cluster
mass scales with richness (see below), $P$; the age of the universe at the beginning
of the calculation, $t_1$; the minimum distance allowed between clusters, $r_{\rm
min}$; the number of time steps, $N$; and the minimum and maximum redshifts used in
selecting the clusters from the catalogue, $z_{\rm min}$ and $z_{\rm max}$.  

The cluster mass, ${\cal M}_2$,
was taken from Abell (1974) and Kent and Gunn (1982) as $2.3\times 10^{15} {\cal
M}_\odot$, and $F$ was arbitrarily set at 0.1.  Richer clusters contain a
proportionally greater amount of light which exceeds the ratio of increase in the
number of their member galaxies, because the luminosity functions of richer clusters
extend to brighter magnitudes (Gudehus 1995).  The amount of mass in richer clusters
increases faster than the rate of luminosity increase because richer clusters possess
higher mass-to-luminosity ratios (Davies, {\it et al.} 1980).  I model cluster mass
as a function of number of galaxies of a given richness by the relation  ${\cal M}_r =
{\cal M}_2 (n_r/n_2)^P$, where $n$ is the average number in a richness class.  An
estimate of $P$ can be obtained from the velocity dispersions in clusters of two
different richnesses.  Taking values for the Coma cluster (Kent and Gunn 1982) and
cluster A1689 (Gudehus 1989a), I derive $P=2.1$.  Based on Larson (1969), the age at
the beginning of the calculation is taken to be $1.0\times 10^9$ y.  The minimum
cluster separation is assumed to be 1 Mpc, but this parameter rarely comes into play.
The Hubble constant is taken to be 70 {\hbox{km\ s$^{-1}$Mpc$^{-1}$}}.
A number of time steps equal to 1000 gives sufficient accuracy, and the redshift
limits are taken as 0 and 0.3.  

A simulation with the above values includes 1009 clusters of which 1 is of richness 5,
10 of richness 4, 61 of richness 3, 131 of richness 2, 257 of richness 1, 358 of
richness 0, and 190 of unknown richness.  The resultant peculiar velocities and 
galactic coordinates of the vectors are shown in Table~\ref{tbl-2}. Increasing $P$
to 2.8 gives better agreement with the observed velocities of the LG and Virgo's
radial peculiar velocity.  The angular deviation of the LG's direction from the MBR
is 27\deg.  The direction for the simulation with $P=2.8$ has been included in
Figure~\ref{fig-3}.  

\begin{table}
\caption{Cluster Peculiar Velocity Simulations.} \label{tbl-2}
\begin{center}
\begin{tabular}{lccccr}
    & $v_{\rm LG}$           &$v_{\rm Virgo}$           &  $v_{r,{\rm Virgo}}$    &
                 long$_{\rm LG}$    & lat$_{\rm LG}$  \\
$P$ & {\hbox{(km\ s$^{-1}$)}} & {\hbox{(km\ s$^{-1}$)}} & {\hbox{(km\ s$^{-1}$)}} & 
                 long$_{\rm Virgo}$ & lat$_{\rm Virgo}$ \\ 
\tableline
2.1      & 600  & 584  & 162 & 304\deg & 23\deg \\
         &      &      &     & 316\deg &  3\deg \\
2.8      & 626  & 712  & 348 & 308\deg & 33\deg \\
         &      &      &     & 317\deg & 17\deg \\
\end{tabular}
\end{center}
\end{table}

\begin{figure}
\plotfiddle{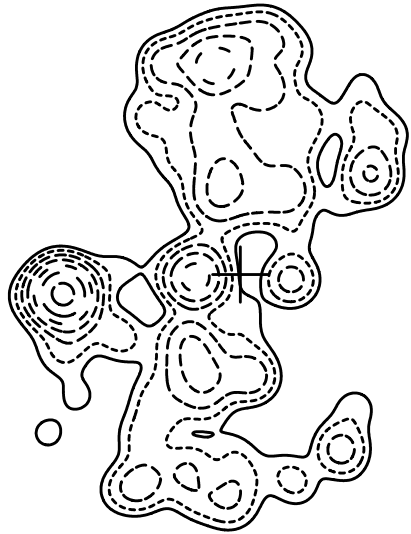}{3.0in}{0}{65}{65}{-200}{-30}
\caption{A contour plot of mass in the N-body model used for
simulating peculiar motions, which is associated with clusters of known redshift
that are  closer than $z=0.07$, and which lie within 45 degrees of the best-fitting
plane through the eight nearby clusters.   The locations, representing 366
clusters, were smoothed with a gaussian of 1232 {\hbox{km\ s$^{-1}$}}. 
 The Local Group is at the center.  The peaks represent extragalactic mass
concentrations (EMACS) made up of very  rich clusters and their associated dark 
matter.   For example, the peak at the left at about $z=0.05$ includes the richness
4 cluster Abell 3558 and the richness 3 cluster Abell 3559.  The Perseus cluster is
the small peak just to the right of center.  The MBR flow is to the left.} 
\label{fig-4}
\end{figure}

\begin{figure}
\plotfiddle{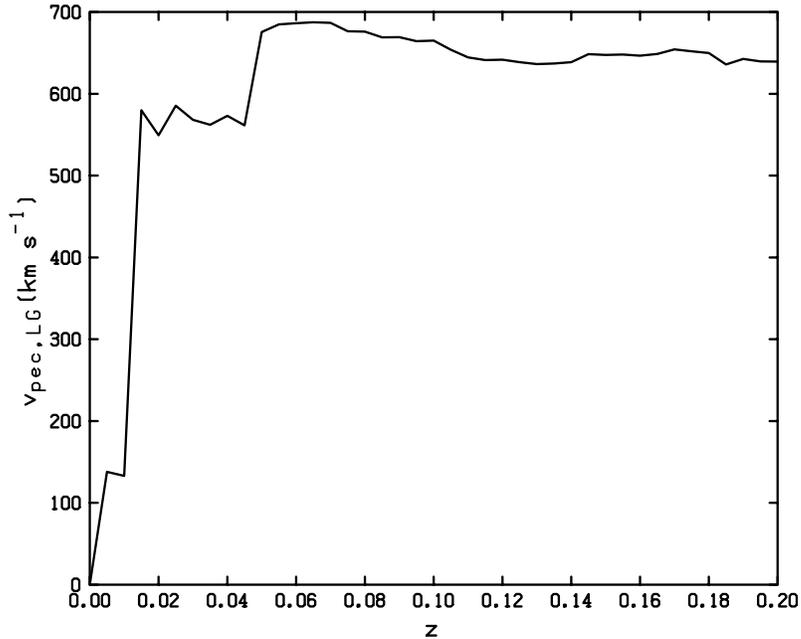}{3.0in}{0}{60}{60}{-190}{-10}
\caption{The magnitude of the peculiar velocity of the Local Group as
a function of the upper redshift limit of the N-body model.  Most of the peculiar
velocity is accounted for by clusters within about 40 $h^{-1}$Mpc ($z=0.013$).  An
additional contribution at 150 $h^{-1}$Mpc ($z=0.05$) arises from clusters which
include Abell 3558 and Abell 3559.  Convergence to a steady value is
reached after about 450 $h^{-1}$Mpc ($z=0.15$).}
\label{fig-5}
\end{figure}

Figure~\ref{fig-4} shows a smoothed false color plot of the mass assigned to the
clusters closer than $z=0.07$ and which lie within 45\deg\ of the best-fitting
plane through the eight nearby clusters.  Since the best fitting $P$ is greater
than the value based on known cluster masses, i.e., 2.1, increasing amounts of mass
are contained outside the richer clusters.  Figure~\ref{fig-5} shows how the model
velocity for the LG depends on upper redshift limit.  Most of the accelerating
masses lie within about 40$h^{-1}$ Mpc ($z=0.013$), but more distant clusters, such
as those at 150$h^{-1}$ Mpc ($z=0.05$) contribute as well.  Convergence to a steady
value occurs only after a distance of about 450$h^{-1}$ Mpc ($z=0.15$) is reached.

\section{Summary}

Both the magnitude and direction of the observed average radial peculiar velocity are
consistent with the set of eight clusters participating in the same flow that the LG
moves with.  Gudehus's (1995) observation that when the measured infall to Virgo
of 52 {\hbox{km\ s$^{-1}$}}  is used to compute the LG's undisturbed velocity vector, the projected 
radial velocity precisely matches Virgo's measured radial peculiar velocity,
demonstrates that at least Virgo participates in the flow.  Other clusters in that
study also showed evidence of flow participation.

Simulations for gravitationally induced peculiar motions in an expanding universe
with model parameters set equal to reasonable values drawn from the literature give
acceptable agreement with the observed velocities of the LG and Virgo.  The sparseness
of the cluster catalogue prevents extending the calculation to more distant clusters
at this time.

\end{document}